\documentclass[twocolumn,prl,showpacs]{revtex4}

\usepackage{graphicx}
\usepackage{amsmath, amssymb}
\usepackage{natbib}

\begin{document}


\title{Spin relaxation in the presence of electron-electron interactions.}

\author{Alexander Punnoose}%
 \email{apunnoose@wisc.edu}
  \affiliation{Lucent Technologies, Bell Labs,  Murray Hill, NJ 07974, USA}
  \altaffiliation[Current Address: ]{Physics Department, University of Wisconsin, 1150 University Ave., Madison, WI 53706}
\author{Alexander M. Finkel'stein}
 \affiliation{Department of Condensed Matter Physics, Weizmann Institute of
Science, Rehovot 76100, Israel}
 \affiliation{Argonne National Laboratory, Argonne, IL 60439, USA}

\begin{abstract}
The D'yakonov-Perel' spin relaxation  induced by the spin-orbit
interaction is examined in disordered two-dimensional electron gas.
It is shown that, because of the electron-electron interactions
different spin relaxation rates can be obtained depending on the
techniques used to extract them.
It is demonstrated that the relaxation rate of a spin population is
proportional to the spin-diffusion constant $D_s$, while the
spin-orbit scattering rate controlling the weak-localization
corrections is proportional to the diffusion constant $D$, i.e., the
conductivity.
The two diffusion constants get strongly renormalized by the
electron-electron interactions, but in different ways.
As a result, the corresponding relaxation rates are different, with
the difference between the two being especially strong near a
magnetic instability or near the metal-insulator transition.
\end{abstract}


\maketitle

Understanding the various mechanisms controlling the rate of spin
relaxation is important for the further development of spintronics.
In disordered systems the dominant spin relaxation mechanism
in the presence of spin-orbit (SO) interactions is the
D'yakonov-Perel' (DP) mechanism~\cite{dp}.
Due to the spin splitting, $\Delta_{so}$, induced by the SO
interaction the electron-spin precesses  through the angle
$\Delta_{so}\tau$ between collisions with the impurities, where
$\tau^{-1}$ is the momentum relaxation rate. When the spin splitting
is weak, $\Delta_{so}\tau \ll 1$,  the precession axis is changed
randomly at each collision. The net precession after $N$ such
collisions is typically $\sqrt{N}\Delta_{so}\tau$ rather than
$N\Delta_{so}\tau$. Consequently, the time taken for the electron
spin to randomize  is $\sim 1/\Delta_{so}^2\tau$.
The spin relaxation  is therefore directly related to the transport
properties of the electrons. Indeed, spin relaxation investigated in
metallic semiconductor systems clearly show that the DP mechanism is
the dominant relaxation mechanism in the metallic phase, with the
relaxation rate decreasing as the metal-insulator transition is
approached; see Ref.~\cite{sandhu01} and especially the extended
discussion of Fig.~3 in Ref.~\cite{dzhioev02}.

It is known that electron-electron ($e\textrm{-}e$) interactions
play an essential role in determining both the transport and the
thermodynamic properties near the metal-insulator transition in
two-dimensional (2d) electron systems~\cite{kravreview04}. In
particular, the spin susceptibility has been shown to behave
critically as the metal-insulator transition is
approached~\cite{mishachi03,kravchi04,punnoose05}, indicating that
the $e\textrm{-}e$ interactions may also have a profound effect on
the rate of spin relaxation. In this paper, we show that because of
the electron-electron interactions the spin relaxation caused by the
DP mechanism gives considerably different rates depending on the
technique employed.

Optical orientation methods are used extensively in semiconductors
to create a spin-polarized population of electrons. The subsequent
relaxation of the spins is then monitored to study both the
temporal~\cite{barad92,samarth94} and the spatial dynamics of the
spin subsystem~\cite{cameron96}.
In the presence of impurities, the spin density propagates
diffusively at low frequencies and large distances. One can find how
the $e\textrm{-}e$ interactions modify this propagation  by studying
the dynamic (retarded) spin susceptibility:
\begin{equation}
\chi_s^{xx}(q,\omega)=i\int_0^{\infty} dt\ e^{i\omega t}\langle\left[S^x(t),S^x(0)\right]\rangle~.
\label{eqn:chidef}
\end{equation}
In the presence of Fermi-liquid corrections it acquires the form:
(we use Matsubara frequencies):
\begin{eqnarray}
\chi_s^{xx}(q,\omega_n)&=&\chi_s^0(1+\Gamma_2)\
\frac{Dq^2+\Delta_{so}^2\tau/2}{(1+\Gamma_2)\;
\omega_n+Dq^2+\Delta_{so}^2\tau/2}\nonumber\\
&=&\chi_s\ \frac{D_s q^2+D_s(\Delta_{so}/v_F)^2}{\omega_n+D_s
q^2+D_s(\Delta_{so}/v_F)^2}~. \label{eqn:FLchi}
\end{eqnarray}
Here, 
$\chi_s=\chi_s^0(1+\Gamma_2)$ is the Pauli spin susceptibility
enhanced by the Stoner factor $(1+\Gamma_2)$~\cite{note:f0a}; for
free electrons $\chi_s^0$ is proportional to the single-particle
density of states~$\nu$.
%
In the presence of a finite SO interaction spin is not conserved
(for the reasons discussed above) and correspondingly
$\chi_s^{xx}(0,\omega)\neq 0$~\cite{arcadi05}.
From Eq.~(\ref{eqn:FLchi}), it follows that the rate of relaxation
of a spin population due to the DP mechanism is:
\begin{equation}
\tau_{{s}}^{-1}=D_s(\Delta_{so}/v_F)^2~, \label{eqn:taudp}
\end{equation}
where $D_s=D/(1+\Gamma_2)$~\cite{sasha84,castellani84} is the
spin-diffusion constant, and $D=v_F^2\tau/2$  is the diffusion
constant of 2d electrons. (We assume that $v_F$ and $\Delta_{so}$
include the standard Fermi-liquid corrections~\cite{raikh99}.)

An alternate way to study the SO scattering time is to study the
effect of SO interaction on the quantum interference correction to
the conductivity~\cite{knap}. The SO scattering rate introduces a
cutoff  $1/\tau_{so}$  in the triplet part of the interference
processes, leading to the well-known anti-localization
effect~\cite{HLN}. In the case of the DP mechanism, the SO rate is:
\begin{equation}
\tau_{so}^{-1}=\Delta_{so}^2\tau/2= D(\Delta_{so}/v_F)^2~.
\label{eqn:tauso}
\end{equation}
Note that, $\tau_{so}^{-1}$ is proportional to $D$, and not $D_s$ as
in Eq.~(\ref{eqn:taudp}).

Since the two rates $\tau_{{s}}^{-1}$  and $\tau_{so}^{-1}$ are
directly related to the  behavior of the different transport
coefficients, let us clarify the relationship between the diffusion
constants in a disordered electron
liquid~\cite{sasha84,castellani84}. The charge-diffusion constant
$D_{c}$ is related to the conductivity $\sigma$ and the diffusion
constant $D$ by the Einstein relation:
\begin{equation}
\sigma/e^2=\frac{\partial n}{\partial \mu}D_{c}=2\nu D~.
\label{eqn:einsteincharge}
\end{equation}
The spin-diffusion constant $D_s$ is in turn related to $D$ by the
``Einstein" relation for the spin density as:
\begin{equation}
\chi_s D_s={\chi_{s}^0}D~. \label{eqn:einsteinspin}
\end{equation}
The compressibility $\partial n /\partial \mu$ controls the static
limit of the polarization operator just like $\chi_s$ controls the
static limit of the spin-spin correlation function.
Eqs.~(\ref{eqn:einsteincharge}) and (\ref{eqn:einsteinspin}) taken
together reflect the fact that both the charge and the spin are
carried by the same particles. Because of the rather different
renormalization of the respective charge and spin-diffusion
constants, a strong deviation in the rates $\tau_{{s}}^{-1}$ and
$\tau_{so}^{-1}$ may occur. The difference between the two rates can
be especially strong near a magnetic instability or near the
metal-insulator transition~\cite{punnoose05} where
$\chi_s\rightarrow\infty$ whereby $D_s\rightarrow 0$  and hence
$\tau_s^{-1}\rightarrow 0$, while both $D$ and $\tau_{so}^{-1}$
remain finite.

In disordered conductors in 2d the parameters of the electron
liquid, in particular the diffusion constants $D$ and $D_s$, acquire
logarithmically divergent corrections as a function of temperature
due to the combined action of the $e\textrm{-}e$ interaction and
disorder~\cite{AAbook}. The question arises how the different spin
relaxation rates are renormalized as a result of these corrections.
%
In this paper we show that for the practically important case of the
Bychkov-Rashba SO interaction the expressions observed in
Eqs.~(\ref{eqn:taudp}) and (\ref{eqn:tauso}) at the Fermi-liquid
level are preserved in the course of the logarithmic
renormalizations.

In 2d systems with structure inversion asymmetry, typical to
heterostructures, the SO interaction has the form of the
Bychkov-Rashba term~\cite{rashba}:
\begin{equation}
H_{so}=\frac{\vec{p}^{\;2}}{2m}+\alpha_{so}~\vec{\sigma}\cdot
(\hat{z}\times\vec{p})~.
\label{eqn:rashba}
\end{equation}
Naively, the SO interaction can be looked upon as a momentum
dependent ``Zeeman" interaction. While the Zeeman splitting induced
by a magnetic field is strongly renormalized in the presence of
disorder and $e\textrm{-}e$
interactions~\cite{sasha84,castellani84}, we find that the momentum
dependence of the SO interaction radically changes the situation.
The momentum dependence allows $H_{so}$ to be rewritten (up to a
constant) in the gauge form:
\begin{equation}
H_{so}=\frac{1}{2m}\left(\vec{p}+{p_{so}}~\vec{\tau}\ \right)^2~,
\label{eqn:rashbavector}
\end{equation}
where, the SO interaction appears as a spin dependent vector
potential, $\vec{\tau}=\frac{1}{2}(\vec{\sigma}\times\hat{z})$, with
effective charge $p_{so}=2m\alpha_{so}$. (In terms of $p_{so}$, the
spin splitting at the Fermi-surface $\Delta_{so}=p_{so}v_F$.)  We
show below  that $p_{so}$, unlike the Zeeman term, is not
renormalized.

The renormalization of the parameters of the disordered electron gas
is best described by the matrix non-linear sigma
model~\cite{wegner,sasha83}. The  SO interaction can also be
succinctly described within this model. The disorder-averaged
N-replica partition function of the interacting problem reads:
\begin{eqnarray}
\langle Z_N\rangle&=&\int DQ~e^{-S[Q]}~,\\
S[Q]&=&\int d^2 r~\frac{\pi \nu}{4\tau}\; \textrm{Tr}\;Q^2 -\textrm{Tr} \ln
G^{-1} +Q\hat\Gamma Q~,
\label{eqn:QG}
\end{eqnarray}
where the $Q$-field is an auxiliary matrix field describing the impurity scattering.
The Green's function in Eq.~(\ref{eqn:QG})
is easily deduced from Eq.~(\ref{eqn:rashbavector}):
\begin{equation}
G^{-1}=i\epsilon+\mu-\frac{1}{2m}(-i\vec{\nabla}+{p_{so}}~\vec{\tau}\
)^2+\frac{i}{2\tau}~Q~,
\end{equation}
with the last term appearing because of the impurity scattering. On
expanding the action $S[Q]$ about its saddle point solution the
problem of electron diffusion in the field of impurities,
$e\textrm{-}e$ interactions, and SO
interaction~\cite{gauge:bleibaum} reduces to the non-linear sigma
model:
\begin{widetext}
\begin{eqnarray}
S[Q]&=&\frac{\pi \nu}{4}\int \textrm {Tr}\; D\left(\vec{\nabla}Q+
i~{p_{so}} \left[\vec{\tau},Q\right]\right)^2 -4z
\textrm{Tr}\left(\hat{\epsilon}Q\right) + Q\hat\Gamma Q~.
\label{eqn:action}
\end{eqnarray}
\end{widetext}
The matrix $Q$ satisfies the constraints:
$Q^2 = 1, Q=Q^\dagger,$ and Tr\;$Q =0$. The components
of $Q$ are defined as $Q^{ij,\alpha\beta}_{n_1,n_2}$, where $n_1, n_2$ are the
fermionic
energy indices with $\epsilon_n=(2n+1)\pi T$; $i,j$ are the replica indices
and $\alpha, \beta$ are the spin indices. The trace is taken
over all these variables. The energy matrix
$\hat{\epsilon}=\epsilon_n\delta_{nm}\delta_{ij}\delta_{\alpha\beta}$.
The factor $z$ is the frequency renormalization factor~\cite{sasha83}
that determines the relative scaling of the frequency
with respect to the length scale; $z=1$ for free electrons.

The fluctuations of the  $Q$-field in the particle-hole channel
(diffusons) determine the propagators $\mathcal{D}(q,\omega)$, which
in the absence of the SO interaction ($p_{so}=0$) have a
diffusion-like singularity $\mathcal{D}(q,\omega)=1/(Dq^2+z\omega)$.
These propagators describe the diffusive evolution of the charge and
spin density fluctuations at large scales. The fluctuations
responsible for the interference corrections are in the
particle-particle channel (cooperons). Since the Bychkov-Rashba SO
interaction is invariant under time-reversal, the particle-hole
diffuson and the particle-particle cooperon propagators are related
by charge conjugation~\cite{mathurAB,edelstein95}. It therefore
suffices to study the renormalization of the spin-orbit scattering
in the diffuson channel only.

The spin fluctuations can be classified in terms of the total spin,
$S$, of the particle-hole pairs. The interaction matrix
$\hat{\Gamma}=\Gamma_s$ and $\Gamma_t$ represent the multiple
scattering induced by the $e\textrm{-}e$ interactions in the singlet
$S=0$ and the triplet $S=1$ diffuson channels, respectively.
They are related to the  static Fermi-liquid parameters $\Gamma_1$
and $\Gamma_2$ as $\Gamma_s=\Gamma_1-{\Gamma_2}/{2}$ and
$\Gamma_t=-{\Gamma_2}/{2}$.
The presence of the quadratic term  $Dp_{so}^2
\left(Q\vec{\tau}Q\vec{\tau}\right)$ in Eq.~(\ref{eqn:action})
introduces in the triplet $S=1$ diffuson and cooperon propagators a
cutoff, i.e., a gap, proportional to the SO scattering rate
$\tau_{so}^{-1}=Dp_{so}^2$.
Note that the linear term proportional to $Dp_{so}(\vec{\nabla}
Q\vec{\tau} Q)$ in Eq.~(\ref{eqn:action}) may be gauged away by the
transformation ${Q}(\vec{r})\rightarrow e^{{i} p_{so}
\vec{\tau}\cdot\vec{r}}Q(\vec{r})$~\cite{gauge:AF}. This will,
however, generate higher order terms in $p_{so}$ since Pauli
matrices do not commute with each other. The question of the
renormalization of the quadratic term therefore still remains.

\begin{figure}
\hspace{-1.5cm}\includegraphics[width=0.75\linewidth]{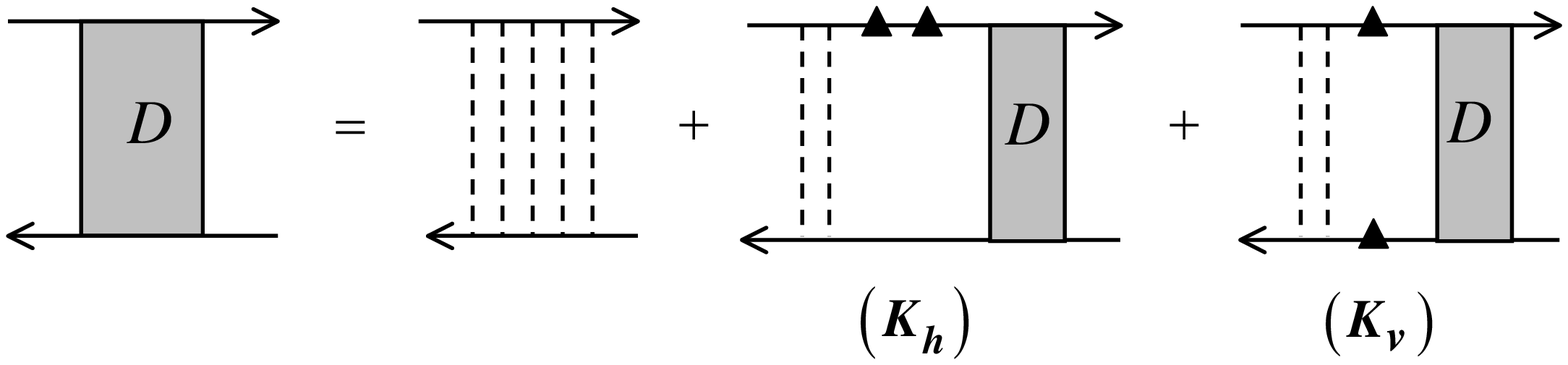}
\caption{Kernels $K_h$ and $K_v$ indicating two possible ways to
insert the spin-orbit vertices $(\blacktriangle)$ into the
propagator $\mathcal{D}$.} \label{fig:bareD}
\end{figure}

The quadratic term leads to inserting two $\vec{\tau}$-vertices
into the particle-hole  propagator $\mathcal{D}$.  These vertices are marked by
the solid triangles ($\blacktriangle$) in Fig.~\ref{fig:bareD}.
As shown
in the figure, there are two possible ways
indicated by the kernels $K_h$ and $K_v$ in which this can be done.
To determine the scaling behavior of the cutoff $\tau_{so}^{-1}$ in
the diffuson channel we study how both these kernels are
renormalized when the  $e\textrm{-}e$ interaction corrections are
included in the presence of disorder.

Since each momentum integration involving the diffusion propagators
leads to one power of $\rho=1/(2\pi)^2\nu D$ in 2d,
the corrections to first order in the disorder strength $\rho$
are evaluated by taking into account diagrams with only one momentum.
In the following, we demonstrate how the scaling equations for
$\tau_{so}^{-1}$ are obtained by considering the renormalization of
the kernel $K_v$ as an example.

\begin{figure}
\vspace{2\baselineskip}
\includegraphics[width=0.4\linewidth]{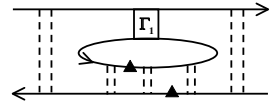}
\caption{ $\Gamma_1$ corrections to the $K_v$ kernel in
Fig.~\ref{fig:bareD} involving only one momentum  integration.}
\label{fig:rggamma1}
\end{figure}

\begin{figure}
\vspace{\baselineskip}
\includegraphics[width=0.7\linewidth]{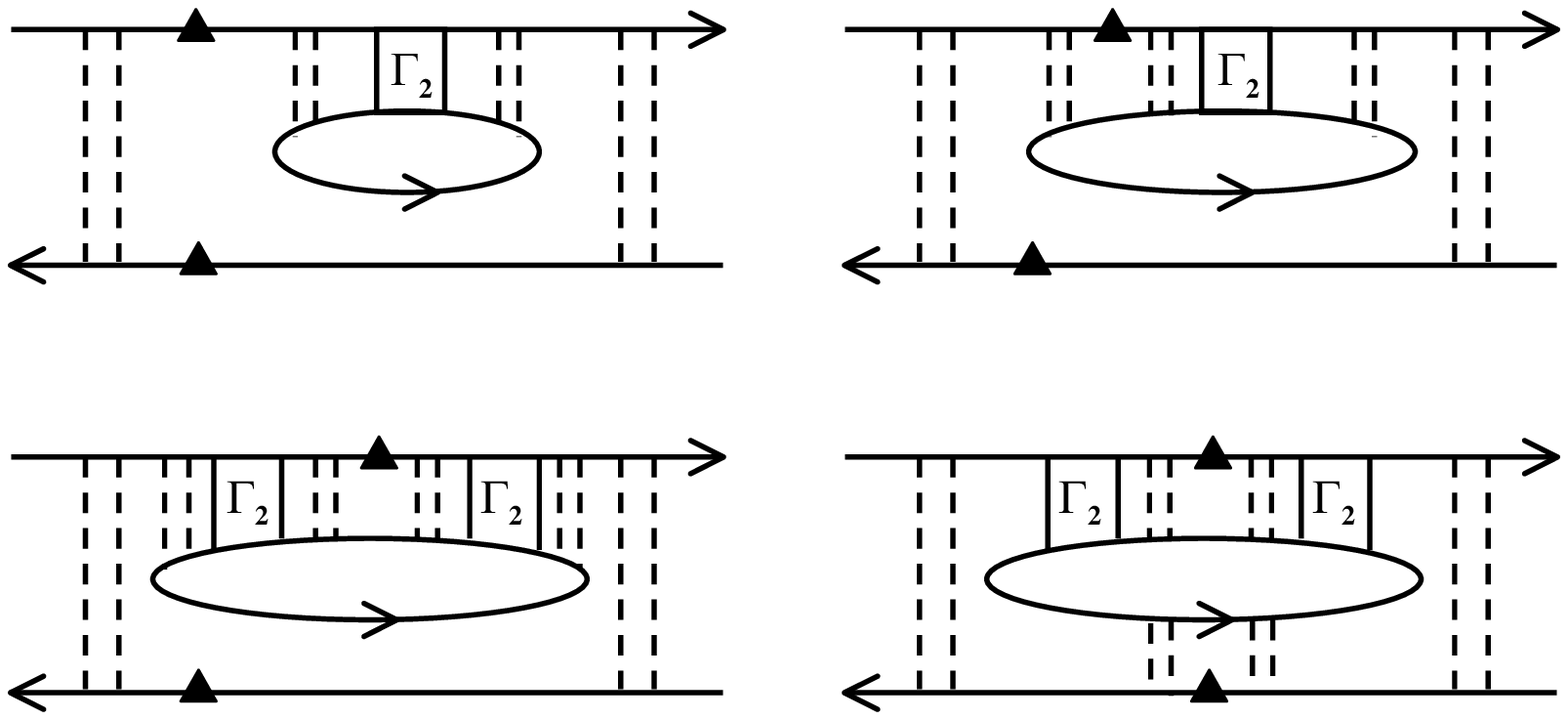}
\caption{ $\Gamma_2$ corrections to the $K_v$ kernel in
Fig.~\ref{fig:bareD} involving only one momentum  integration.}
\label{fig:rggamma2}
\end{figure}

Corrections to  $K_v$ limited to only one momentum integration
involving the interactions  $\Gamma_1$ and $\Gamma_2$ are shown in
Figs.~\ref{fig:rggamma1} and \ref{fig:rggamma2}, respectively.
(Similar corrections exist for the $K_h$ kernel as well.) With the
exception of the first diagram in Fig.~\ref{fig:rggamma2}, these
corrections originate from the linear SO term
$iDp_{so}(\vec{\nabla}Q\vec{\tau}Q)$ applied twice.
They lead to the renormalization of the cutoff  $1/\tau_{so}$. The
renormalized cutoff  $1/\tau_{so}'$ can be expressed in terms of the
renormalized parameters $D\rightarrow D'$ and $p_{so}\rightarrow
p_{so}'$:
\begin{equation}
1/\tau_{so}'= D'{p_{so}'}^2=Dp_{so}^2 \left(1+\frac{\delta
D}{D}+2\frac{\delta p_{so}}{p_{so}}\right)~,
\end{equation}
where $\delta D$ and $\delta p_{so}$ are the corrections to $D$ and
$p_{so}$, respectively. Given $\delta D$, one can extract $\delta
p_{so}$ from the calculated $1/\tau_{so}'$. $\delta D$ corresponds
to  the well known Altshuler-Aronov corrections to the
conductivity~\cite{AAbook}:
\begin{equation}
\frac{\delta D}{D}=-\frac{4}{\nu}\int \frac{d^2q}{(2\pi)^2}\int\frac{d\omega}{2\pi}\;
\mathcal{D}^3(q,\omega) Dq^2(\Gamma_1-2\Gamma_2)~.\label{eqn:deltaD}
\end{equation}
Then for  $\delta p_{so}$, we get:
\begin{widetext}
\begin{equation}
\frac{\delta p_{so}}{p_{so}}=\frac{4}{\nu}\int \frac{d^2q}{(2\pi)^2}\int\frac{d\omega}{2\pi}\;
\biggl[\mathcal{D}^2(q,\omega)\Gamma_2-2\mathcal{D}^3(q,\omega)Dq^2\Gamma_2
+\mathcal{D}^4(q,\omega)\omega Dq^2\Gamma_2^2 \biggr]~.\label{eqn:deltap}
\end{equation}
\end{widetext}
Note that $\delta p_{so}$ depends only on the triplet amplitude.

The corrections given in Eqs.~(\ref{eqn:deltaD}) and
(\ref{eqn:deltap}) have so far employed the static amplitudes
$\Gamma_1$ and $\Gamma_2$. To include the effects of dynamical
screening, these amplitudes are extended via the ladder
summation~\cite{sasha83}:
$U_{s,t}(q,\omega)=\Gamma_{s,t}
({Dq^2+z\omega})/({Dq^2+(z-2\Gamma_{s,t})\omega})$.
The dynamical re-summation  allows the evaluation of the corrections
to infinite order in the interaction amplitudes so that $\rho$
remains as  the only expansion parameter. Remarkably, $\delta
p_{so}$ vanishes after this replacement of the $\Gamma_2$ amplitude
in Eq.~(\ref{eqn:deltap}) by $U_t(q,\omega)$:
\begin{equation}
\delta p_{so}=0~.
\end{equation}

Similar analysis when extended to the kernel $K_h$ and to the linear
term  $Dp_{so}(\vec{\nabla}Q\vec{\tau}Q)$ in Eq.~(\ref{eqn:action})
leads to the result that the gauge form $ D(\vec{\nabla}
Q+{i}p_{so}[\vec{\tau},Q])^2$ of the action in
Eq.~(\ref{eqn:action}) with the bare value of $p_{so}$  is preserved
under renormalization. This implies that the cutoff
$\tau_{so}^{-1}=Dp_{so}^2$ retains the same form as in
Eq.~(\ref{eqn:tauso}) with $D$ as the renormalized diffusion
constant and with $p_{so}$  unrenormalized.

Note that the diagrams presented in Figs.~\ref{fig:rggamma1} and
\ref{fig:rggamma2} resemble those that are used for the calculation
of the spin susceptibility~\cite{sasha84,castellani84} with the
triangle at the top as the starting point and ending at the bottom
triangle. This analogy is, however, misleading since the SO
interaction $Dp_{so}(\vec{\nabla}Q\vec{\tau}Q)$ in addition to the
Pauli matrices is proportional to momentum. It is this dependence on
momentum that makes the calculation of the renormalization of
$p_{so}$ different from the renormalization of the Zeeman term.

We now study the renormalization effects on the DP spin relaxation
rate $\tau_{{s}}^{-1}$. To this end, we analyze the form of the
renormalized dynamic spin susceptibility that follows from the
action defined in  Eq.~(\ref{eqn:action}):
\begin{eqnarray}
\chi_s^{xx}(q,\omega_n)&=&\chi_s^0(z+\Gamma_2)\
\frac{Dq^2+Dp_{so}^2}{(z+\Gamma_2)\;
\omega_n+Dq^2+Dp_{so}^2}\nonumber\\
&=&\chi_s\ \frac{D_s q^2+D_s p_{so}^2}{\omega_n+D_s q^2+D_s
p_{so}^2}~. \label{eqn:DFLchi}
\end{eqnarray}
where, the renormalized spin-diffusion constant:
\begin{equation}
D_s=D/(z+\Gamma_2)~.
\end{equation}
Note that Eq.~(\ref{eqn:DFLchi}) has the same structure as the
expression for $\chi_s^{xx}(q,\omega)$ in the disordered
Fermi-liquid with the parameter $(z+\Gamma_2)$ substituted for
$(1+\Gamma_2)$ in $\chi_s$ and $D_s$. It follows from
Eq.~(\ref{eqn:DFLchi}) that the form of the DP spin relaxation rate,
$\tau_{{s}}^{-1}=D_s p_{so}^2$, given in Eq.~(\ref{eqn:taudp}) is
also preserved under renormalization.

To conclude, we have shown that the form of the functional $S[Q]$ in
Eq.~(\ref{eqn:action}) is preserved under renormalization by  the
combined action of the $e\textrm{-}e$ interaction and disorder. It
follows from this observation that  the relaxation  rate of the spin
density and that of the cutoff in the $S=1$ cooperon/diffuson
propagators are given by Eqs.~(\ref{eqn:taudp}) and
(\ref{eqn:tauso}), respectively, where the parameters  $D_s$ and $D$
are the renormalized diffusion constants~\cite{note:weber}, and the
spin splitting $p_{so}=\Delta_{so}/v_F$ is unchanged. 
As a result, under the circumstances when $\chi_s$ diverges, either
near a magnetic instability ($\Gamma_2\rightarrow\infty$), or near
the metal-insulator transition
($z\rightarrow\infty$)~\cite{punnoose05}, the spin-diffusion
constant $D_s$ and $\tau_{{s}}^{-1}$ vanish, while the scattering
rate  $\tau_{so}^{-1}$ remains finite.

\acknowledgments AF thanks Y. Ji and  Ora Entin-Wohlman for valuable
remarks. AF is supported by the  MINERVA Foundation and by U.S.
Department of Energy under contract W-31-109-ENG-38.

\bibliographystyle{apsrev}

\end{document}